\documentclass{ifacconf}

\usepackage{graphicx}      
\usepackage{natbib}        
\usepackage{amssymb,amsmath,amsfonts}
\usepackage{algorithmic}
\usepackage{mathdots}
\usepackage{lipsum}
\usepackage{floatrow}
\usepackage{threeparttable}
\usepackage{multirow}
\usepackage{comment}
\usepackage{booktabs}

\begin{document}
\begin{frontmatter}

\title{Modeling and Constraint-Aware Control of Pressure Dynamics in Water Electrolysis Systems\thanksref{footnoteinfo}} 

\thanks[footnoteinfo]{This material is based upon work supported by the U.S. Department of Energy’s Office of Energy Efficiency and Renewable Energy (EERE) under the Solar Energy Technologies Office Award Number DE-EE0010147. The views expressed herein do not necessarily represent the views of the U.S.Department of Energy or the United States Government}

\author[First]{Mostafaali Ayubirad} 
\author[First]{Madiha Akbar} 
\author[First]{Hamid R. Ossareh}

\address[First]{Department of Electrical and Biomedical Engineering, Univ. of Vermont, Burlington, VT, 05405 USA (e-mail: {mayubira, Madiha.Akbar,hamid.ossareh} @uvm.edu).}

\begin{abstract}                
This paper addresses the challenge of pressure constraint violations in water electrolysis systems operating under dynamic power conditions, a problem common to both Proton Exchange Membrane and alkaline technologies. To investigate this issue, a control-oriented model of an alkaline electrolyzer is developed, capturing key pressure and flow dynamics. To manage rapid power fluctuations that may cause pressure to exceed manufacturer-defined operational boundaries, a model-based constraint-aware power governor based on the Reference Governor (RG) framework is proposed. Simulation results show that the strategy effectively maintains pressure within the specified operating range, outperforming conventional filtering methods while enhancing hydrogen production and reducing auxiliary energy consumption.
\end{abstract}


\end{frontmatter}

\section{Introduction}
As the global energy landscape evolves, the demand for efficient and scalable energy storage solutions is increasing. Hydrogen has recently emerged as a promising energy carrier due to its high energy density and notable calorific value, making it well-suited for compact, large-scale, and long-duration storage applications (\cite{akyuz2024hydrogen}). Among the available hydrogen production methods, water electrolysis is currently regarded as the most efficient technology, known for producing high-purity and scalable green hydrogen, particularly when powered by renewable sources (\cite{akyuz2024hydrogen}). 
Efforts to reduce the cost of green hydrogen have focused on several electrolysis methods, such as alkaline, proton exchange membrane (PEM), solid oxide (SO), and anion exchange membrane (AEM) systems. Alkaline and PEM electrolysis, in particular, have reached higher levels of technological readiness and commercialization, driving rapid progress in their real-world application (\cite{taibi2020green}).
However, despite the advancements, large-scale deployment of electrolyzers in energy infrastructure remains limited by persisting safety, durability, and cost challenges (\cite{hu2022comprehensive}). Therefore, to tackle these challenges, gaining a deeper understanding of the electrolyzers and developing efficient control mechanisms for improving their operational flexibility is crucial. For these reasons, in recent years, a significant effort has
been made to model, characterize and analyze the operation of electrolyzers.

In this regard, one of the most widely used models to describe the behavior of both PEM and alkaline electrolyzers was developed by \cite{ulleberg2003modeling}. The model describes the polarization curve and also accounts for the thermal behavior of the system and the Faraday efficiency of the stack. To improve voltage prediction, \cite{sanchez2018semi} modified Ulleberg’s model to account for pressure effects in the polarization curve.
\cite{david2020dynamic} and \cite{cantisani2023dynamic} developed dynamic models of electrolyzer pressure based on mass balance and thermodynamic principles, providing accurate representations of system pressure. However, these large-scale, physics-based models involve dozens of parameters that must be tuned experimentally through a complex identification process. This highlights the need for a simpler, yet effective, control-oriented model that can capture the essential pressure dynamics for the development of pressure control strategies.

Pressure control is a key objective for ensuring the safe and efficient operation of electrolyzer systems. Existing studies on electrolyzer pressure control typically rely on methods such as Proportional-Integral (PI) to maintain constant system pressure, often by treating the applied power as a disturbance to be rejected (\cite{aguirre2024control}).  
Severe power ramps can jeopardize the operation of the electrolyzer (\cite{li2023exploration}), as the system pressure may exceed the acceptable range defined by the manufacturer during transient conditions.
To mitigate these issues and extend the electrolyzer’s lifespan, a controlled, gradual change in power input using approaches such as a conventional low-pass filter (LPF) has been recommended during instantaneous transients (\cite{ zhou2009modeling, li2023exploration}).  Low-pass filtering of the input power can help prevent constraint violations but is usually  conservative. Currently, the development of constraint-based control strategies for managing changes in input power remains unexplored in the literature. Reference Governor (RG) strategies have been successfully applied in related energy systems, such as hydrogen fuel cell vehicles, to enforce operational constraints in a computationally efficient manner (\cite{ayubirad2024machine,ayubirad2025neural,bacher2023hierarchical}). 
In this paper, we treat the input electric power as a controllable variable and develop a Power Governor (PG) strategy based on the RG framework, which modifies the input power, only as needed, for constraint enforcement.

In sum, the main objective of this research is to investigate a constraint-aware PG for managing pressure in water electrolysis systems under dynamic power input, addressing limitations of existing approaches. As such, the original contributions of this paper are as follows:
\begin{itemize}
    \item A physics-based, control-oriented model of electrolyzer pressure dynamics, including control loops and actuators, that can capture hydrogen pressure violations during large, rapid input power transients.
    \item A constraint enforcement strategy for hydrogen pressure using an RG that ensures pressure remains within acceptable limits by adjusting the input power only when necessary.
    \item A performance comparison between the proposed PG and a conventional LPF, to demonstrate improvements in hydrogen production, power tracking, and auxiliary energy consumption.
\end{itemize}
The remainder of the paper is organized as follows. Section~\ref{sec:model} presents the mathematical modeling of system pressure along with the problem statement. Section~\ref{sec:pressure_RG} introduces the proposed constraint management scheme and presents simulation studies and discussions. Conclusions are provided in Section~\ref{sec:conclusion}.

\section{Control design model and problem description}
\label{sec:model}
This section presents a control-oriented model of the electrolyzer, focusing on pressure dynamics and the motivation for advanced pressure control under power transients. While the developed strategies apply to various types of electrolyzers, we focus on alkaline systems due to their widespread use and to maintain precision in our analysis.

\subsection{Control-Oriented Model}

This section presents a model of the alkaline electrolyzer system. The stack temperature is assumed to remain constant and uniformly distributed at $T_{\text{el}} = 80\,^\circ\text{C}$, as its dynamics are considerably slower than other dynamics considered. To concentrate on hydrogen pressure management, the model emphasizes pressure dynamics, control loops, and actuator behavior.

\begin{figure}
\begin{center}
\includegraphics[width=9.1cm]{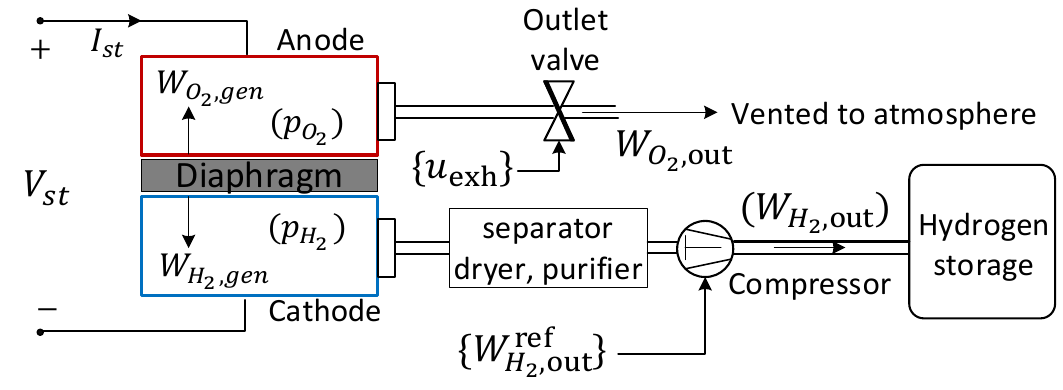}  
\caption{Illustrative electrolyzer system diagram with key variables. The variables in parentheses are state variables, while the variables in braces are input variables determined by the controllers.} 
\label{fig:elec}
\end{center}
\end{figure}

The dynamic pressures of reaction products inside the electrolyzer  are computed based on the cathode and anode flow models. The main flows associated with the electrolyzer stack are shown in Fig.~\ref{fig:elec}, where $W_{H_2,\text{gen}}$ and $W_{O_2,\text{gen}}$ represent the rates at which hydrogen and oxygen are generated, while $W_{H_2,\text{out}}$ and $W_{O_2,\text{out}}$ denote the flow rates of hydrogen and oxygen leaving the system. 
The mass continuity is used to balance the mass of hydrogen and oxygen, as shown by:
\begin{equation}
\begin{aligned}
\frac{d p_{O_2}}{d t} &= \frac{R_{O_2} T_{\text{el}}}{V_{\text{an}}} \left( W_{O_2,\text{gen}} - W_{O_2,\text{out}} \right), \\
\frac{d p_{H_2}}{d t} &= \frac{R_{H_2} T_{\text{el}}}{V_{\text{ca}}} \left( W_{H_2,\text{gen}} - W_{H_2,\text{out}} \right).
\end{aligned}
\label{eqn:p_H2_p_O2}
\end{equation}
Here, \( p_{O_2} \) and \( p_{H_2} \) denote the partial pressures of oxygen and hydrogen, respectively; \( R_{O_2} \) and \( R_{H_2} \) are the specific gas constants for oxygen and hydrogen; and \( V_{\text{ca}} \) and \( V_{\text{an}} \) represent the cathode and anode volumes, respectively. The thermodynamic constants and fixed model parameters are summarized in Table~A.1 of the Appendix.

\begin{figure}[b!]
\begin{center}
\includegraphics[width=8.0cm]{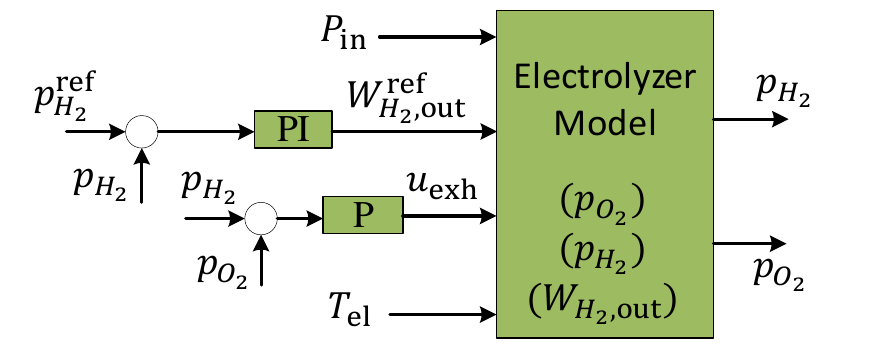}  
\caption{Block diagram of the pressure controllers. The variables in parentheses are state variables.} 
\label{fig:PI_P}
\end{center}
\end{figure}

The layout of the electrolyzer pressure controllers in this paper is shown in Fig.~\ref{fig:PI_P}. 
On the cathode side, the setpoint for the hydrogen outlet flow rate, $W_{H_2,\text{out}}^{\text{ref}}$, is computed by a PI controller based on the error between the actual hydrogen pressure $p_{H_2}$ and its desired value. On the anode side, the control input to the outlet valve, \( u_{\text{exh}} \in [0, 1] \), which determines the degree of valve opening, is computed by a proportional controller based on the difference between the hydrogen pressure \( p_{H_2} \) and the oxygen pressure \( p_{O_2} \) to equalize them. More details about the control-loops and actuators are provided below.

On the cathode side, the hydrogen outlet flow rate is controlled through a compressor, which is assumed to be located downstream of the cathode, enabling the outlet flow rate $W_{H_2,\text{out}}$ to track the reference signal $W_{H_2,\text{out}}^{\text{ref}}$ (see Fig.~\ref{fig:elec}). The hydrogen outlet flow rate is limited by the dynamics of the downstreamed components such as separators, purifiers, dryers, and controlled compressor. Experimental studies have shown that this behavior can be approximated by a simple first-order system (\cite{garcia2012simple}). Therefore, to capture the transient behavior of the hydrogen outlet flow rate in our model, we adopt the following representation, which includes the controlled (i.e. closed-loop) compressor dynamics:

\begin{equation}
\frac{dW_{H_2,\text{out}}}{dt} = \frac{W_{H_2,\text{out}}^{\text{ref}} - W_{H_2,\text{out}}}{\tau_{\text{ca}}}
\label{eqn:W_H2},
\end{equation}
where \( \tau_{\text{ca}} \) is the time constant associated with the hydrogen outlet flow dynamics on the cathode side.
The reference flow rate \( W^{\text{ref}}_{H_2,\text{out}} \) used in \eqref{eqn:W_H2} is determined by the following controller dynamics:
\begin{equation}
\begin{aligned}
\dot{q} &= e_{p_{H_2}}, \\
W_{H_2,\text{out}}^{\text{ref}} &= K_{p,ca} e_{p_{H_2}} + K_i q,
\end{aligned}
\label{eqn:PI_control}
\end{equation}
where \( e_{p_{H_2}} = p_{H_2}^{\text{ref}} - p_{H_2} \) is the hydrogen pressure regulation error, \( q \) is the integrator state, and \( K_{p,ca} \) and \( K_i \) are the constant gains of the PI controller. 


On the anode side of the stack, oxygen would normally undergo similar downstream processing. However, this aspect is not addressed in this paper; instead, the generated oxygen is assumed to be vented, following the approach in \cite{sakas2022dynamic,sanchez2018semi}. As part of the system configuration in alkaline electrolyzers, gas-liquid separators are used on both the anode and cathode sides to collect electrolyte mixed with gas bubbles. These separators are interconnected by a pipe that enables electrolyte transfer between them, helping to balance the pressure across the stack diaphragm and prevent damage. The outlet valve located at the cathode side maintains equal electrolyte levels in the separators (\cite{sanchez2018semi}). To simplify this process in the simulation, the valve is assumed to minimize the pressure difference between both sides. With this assumption, the outlet valve is modeled as a solenoid valve operating under choked flow conditions, where the exhausted oxygen flow rate to the atmosphere is determined by the pressure difference between the two sides of the electrolyzer system and is calculated as follows:
\begin{equation}
W_{O_2,\text{out}} = u_{\text{exh}} \frac{C_D A_T p_{O_2}}{\sqrt{R_{O_2} T_{\text{el}}}}  \gamma^{1/2} \left( \frac{2}{\gamma + 1} \right)^{\frac{\gamma + 1}{2(\gamma - 1)}},
\end{equation}
where \( u_{\text{exh}} = K_{p,\text{an}} (p_{H_2} - p_{O_2}) \), $K_{p,an}$ is the constant gain of the proportional controller on the anode side, \( C_D \) and \( A_T \) are the discharge coefficient and the outlet valve area, respectively, and \( \gamma \) is the ratio of specific heats for oxygen. 

The final components to consider in this modeling are the hydrogen and oxygen production rates in \eqref{eqn:p_H2_p_O2}, which depend on the stack current (\( I_{\text{st}} \), shown in Fig.~1). The stack current is related to the input power of the electrolyzer through \( P_{\text{in}} = V_{\text{st}} I_{\text{st}} \), where \( V_{\text{st}} \) is the stack input voltage. In order to compute the stack voltage and current for a given input power, the polarization curve of the electrolyzer must be incorporated into the model. In this paper, we use the empirical model proposed by \cite{ulleberg2003modeling}, which has been widely utilized by researchers in the field: \( V_{\text{st}} = n_{\text{cell}}V_{\text{cell}}    \), where
\begin{equation}
\begin{aligned}
V_{\text{cell}} =\ & V_{\text{rev}} + (r_1 + r_2 T_{\text{el}})\, i + (s_1 + s_2 T_{\text{el}} + s_3 T_{\text{el}}^2) \times \\
& \log\left( \left( t_1 + \frac{t_2}{T_{\text{el}}} + \frac{t_3}{T_{\text{el}}^2} \right) i + 1 \right).
\end{aligned}
\label{eqn:empirical}
\end{equation}
Here, $V_{\text{cell}}$ is the cell voltage, \( n_{\text{cell}}  \) is the number of cells, \( i = I_{\text{st}} / A_{\text{cell}} \) is the current density, \mbox{\( A_{\text{cell}}   \)} is the surface area of the cell, and \( r_1 \), \( r_2 \), \( s_1 \), \( s_2 \), \( s_3 \), \( t_1 \), \( t_2 \), and \( t_3 \) are empirical parameters taken from Table~1 in \cite{zhou2009modeling}.
The reversible voltage \( V_{\text{rev}} \) is empirically obtained as:
\begin{equation}
V_\text{rev} = 1.518 - 1.542 \times 10^{-3} T_{\text{el}} + 9.523 \times 10^{-5} T_{\text{el}} \ln(T_{\text{el}}). 
\end{equation}

For a given input power, $P_{\text{in}}$, the above equations are used to numerically solve for $i$, and hence, for $V_{\text{st}}$ and $I_{\text{st}}$.
With \( I_{\text{st}} \) computed as such, the hydrogen and oxygen production rates are given by:
\begin{equation}
\begin{aligned}
W_{H_2,\text{gen}} &= \frac{\eta_F n_{\text{cell}} I_{\text{st}}}{2F}, \quad
W_{O_2,\text{gen}} &= \frac{\eta_F n_{\text{cell}} I_{\text{st}}}{4F},
\end{aligned}
\label{eqn:gas_production}
\end{equation}
where \( F \) is the Faraday constant. The parameter \( \eta_F \) in \eqref{eqn:gas_production} is known as the Faraday efficiency, which is defined as the ratio between the actual volume of gas produced and the theoretical volume that should be produced. Several empirical models have been developed in the literature to describe the Faraday efficiency. In this paper, we use the model from \cite{zhou2009modeling}, given as:
\begin{equation}
\eta_{F} = a_1 \exp\left( \frac{a_2 + a_3 T_{\text{el}}}{i} + \frac{a_4 + a_5 T_{\text{el}}}{i^2} \right),
\label{eqn:eta_F}
\end{equation}
where \( a_1 \), \( a_2 \), \( a_3 \), \( a_4 \), and \( a_5 \) are empirical parameters taken from Table~1 in \cite{zhou2009modeling}. The specific parameter values used for numerical simulations in this paper are provided in Table~A.1 of the Appendix.

\subsection{Problem description}
\begin{figure}
\begin{center}
\includegraphics[width=8.4cm]{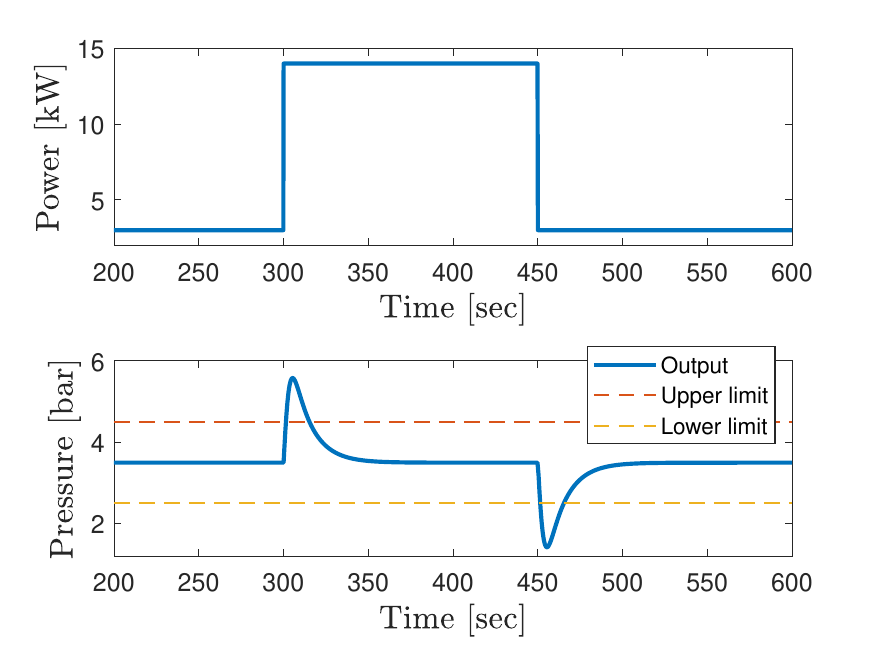} 
\caption{Simulation result of pressure response $P_{H_2}(t)$ under large step changes in input power $P_{\text{in}}(t)$.} 
\label{fig:pressure_management}
\end{center}
\end{figure}

Electrolyzers may have difficulty withstanding severe power ramping (\cite{li2023exploration}). This is because when the power variation is faster than the gas delivery control capacity, the hydrogen pressure can exceed the acceptable range defined by the manufacturer. Fig. ~\ref{fig:pressure_management} illustrates this pressure management issue during a simulation in which the developed model is subjected to a varying input power between 3 and 14~kW. In this scenario, the electrolyzer attempts to maintain a constant pressure of 3.5~bar, while the acceptable range is set between 2.5 and 4.5~bar.
The figure reveals that the system pressure limit is violated during large, rapid changes in input power. Since such changes lead to constraint violations, this paper employs a Power Governor (PG) strategy to manage the applied power to the electrolyzer and maintain the system pressure within its acceptable range.

\section{Pressure Constraint Management}
\label{sec:pressure_RG}
This section focuses on developing a computationally efficient Reference Governor (RG) strategy for pressure management. The model introduced in Section~\ref{sec:model} is employed to maintain the system pressure within the design-specified range defined by the manufacturer of the electrolyzer. The utility of the proposed strategy is evaluated using the plant model developed in the previous section. We begin with a review of the RG approach.
\subsection{Review of Reference Governors}
\label{sec:RG_review}
\begin{figure}[b!]
\begin{center}
\includegraphics[width=8.4cm]{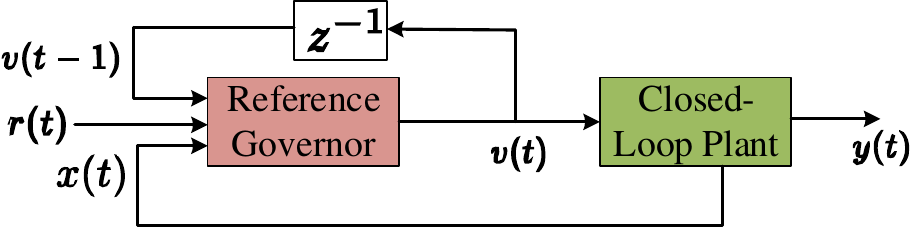}  
\caption{Reference governor controller scheme. The signals are as follows: $y(t)$ is the constrained output, $r(t)$ is the desired reference, $v(t)$ is the modified reference command, and $x(t)$ is the system state.} 
\label{fig:RG}
\end{center}
\end{figure}

Consider Fig.~\ref{fig:RG}, where the “Closed-loop Plant” denotes a discrete-time LTI system in feedback with a stabilizing controller, with closed-loop dynamics described by:
\begin{flalign}
    \begin{split}
        x(t+1) &= Ax(t)+Bv(t), \\ 
        y(t) &= Cx(t)+Dv(t), 
    \end{split}
\label{eqn:LTI}
\end{flalign}
\noindent where $x(t)\in {\mathbb{R}}^{n}$ is the state vector, $v(t)\in \mathbb{R}$ is the modified reference command, and $y(t)\in {\mathbb{R}}^{m}$ is the constrained output. 
Constraints are imposed on the output variables as $y(t) \leq 0$ for all $t \in \mathbb{Z}_+$, where $\mathbb{Z}_+$ denotes the set of all nonnegative integers. The RG approach computes a modified reference command which, if held constant, guarantees constraint satisfaction for all future times. More formally, the output response from the initial state $x_0$ and with a constant input $v_0$ will satisfy the constraints for all future times if the pair $(x_0, v_0)$ lies within the set ${O}_\infty$, known as the maximal output admissible set (MAS), and defined as:
\begin{align}
    O_{\infty} = \left\{ (x_0, v_0) : x(0) = x_0, \ v(j) = v_0, \right. & \nonumber \\
    \left. \Rightarrow y(j) \leq 0, \ j \in \mathbb{Z}_+ \right\},
    \label{eqn:O_inf}
\end{align}
where, under the assumption of constant input, for system~\eqref{eqn:LTI}, ${y}(j)$ is defined as:
\begin{align}
    y(j)=CA^jx_0+[C(I-A^{-1})(I-A^{j})B+D]v_0.
    \label{eqn:y(j)}
\end{align}
The set ${O}_\infty$   require an infinite number of linear inequalities for an exact characterization, which is impossible to compute in finite time. 

\cite{gilbert1991linear} shows that if $A$ is Schur (i.e., all eigenvalues lie strictly within the unit disk) and the pair $(A, C)$ is observable, then, under a tightened steady-state constraint on the output, i.e., $y(\infty) \leq  - \varepsilon$, where $\varepsilon > 0$, there exists a $j^*$ such that $y(j) \leq 0$ for all $j = 0$ to $j^*$ implies $y(j) \leq 0$ for all $j > j^*$. This implies that a computable inner approximation of ${O}_\infty$, which is finitely determined can be constructed as:
\begin{flalign}
    \begin{split}
        \mathrm{\Omega }=\left\{\left(x_0,v_0\right)\ :\ {x}(0)=x_0,\ \ v(j)=v_0,\ \right. &\\
        \Rightarrow  y(\infty)\le -\varepsilon \ ,\ {y}(j) \le 0\ ,\ j=0,\dots ,j^*\left.\right\},
    \end{split}
    \label{eqn:omega}
\end{flalign}
where $\Omega$ can be made arbitrary close to \( O_{\infty} \) by decreasing $\varepsilon $ at the expense of increasing the complexity of $\Omega$. From \eqref{eqn:y(j)} and \eqref{eqn:omega}, $\Omega$ can be explicitly characterized as:
\begin{flalign}
\Omega = \left\{ (x_0, v_0) \;\middle|\; H_x x_0 + H_v v_0 \leq h \right\},
\label{eqn:omega_e}
\end{flalign}
where
\[H_x=\left[ \begin{array}{c}
0 \\ 
C \\ 
CA \\ 
\vdots  \\ 
CA^{j^*} \end{array}
\right],   
H_v=\left[ \begin{array}{c}
{C\left (I-A\right)}^{-1}B+D \\ 
D \\ 
C\left(I-A\right){\left(I-A\right)}^{-1}B+D\ \ \  \\ 
\vdots  \\ 
C\left(I-A^{j^*}\right){\left(I-A\right)}^{-1}B+D \end{array}
\right],
  \] 
and \( h = \left[ -\varepsilon^\top,\, 0^\top,\, \ldots,\, 0^\top \right]^\top \).

The “Reference Governor” in Fig.~\ref{fig:RG} computes $v(t)$ based on $\Omega$ by implementing the following dynamic equation:
\begin{align}
    v(t)=v(t-1)+\kappa(t)(r(t)-v(t-1)),
    \label{eqn:v(t)}
\end{align}
where $r$ is the desired command, and $\kappa(t)\in[0,1]$ is a scalar adjustable bandwidth parameter and is obtained by solving the following Linear Program (LP):
\begin{flalign} 
    \begin{aligned}
        \kappa (t)={\mathop{\max }\limits_{\kappa \in [0,1]}} {\rm \; }\kappa\\
        \text{s.t.} \quad &v=v(t-1)+\kappa \left(r(t)-v(t-1)\right)\\ 
        &(x(t),v)\in \Omega     
    \end{aligned}
\label{eqn:kappa}
\end{flalign}
where we have assumed that the system state $x(t)$ is available at each timestep. The LP problem in \eqref{eqn:kappa} is based on the idea of computing, at each time step $t$ and on the basis of the current state $x(t)$, the optimal control input $v(t)$ that is as close as possible to $r(t)$ while ensuring constraint satisfaction, i.e., $(x(t), v(t)) \in \Omega$. For more information, please refer to \cite{garone2017reference}. 
\subsection{Pressure Management Power Governor (PG)}
In this section, we employ the RG for pressure management of the electrolyzer. In this context, the RG is used as a PG and serves as an interface between the input power and the applied power; that is, it modifies the input power \( P_{\text{in}} \) in Fig.~\ref{fig:PI_P} to a new command signal \( P_{\text{in,mod}} \), which is then applied to the system.
Its role is to adjust the input power in a way that satisfies the pressure constraint, while keeping the modification from the original input as small as possible.

The implementation of the RG discussed in Section~\ref{sec:RG_review} requires a linearized closed-loop model of the electrolyzer. This model is obtained by applying the LTI system analysis in the \textsc{MATLAB}\textsuperscript{\textregistered}/\textsc{Simulink}\textsuperscript{\textregistered} Control System Toolbox to the closed-loop nonlinear system developed in Section~2. We chose the nominal stack power input as \( P_{\text{in}}^o = 7\,\text{kW} \), which corresponds to an input current of \( I_{\text{st}}^o = 177.8\,\text{A} \) and a voltage of \( V_{\text{st}}^o = 39.36\,\text{V} \). The nominal hydrogen pressure is set to \( p_{H_2}^o = 3.5\,\text{bar} \), with the corresponding hydrogen outlet flow rate of \( {W_{H_2,\text{out}}^o = 1.51\,\text{Nm}^3/\text{h}} \) under these conditions. The linearized model is given by:
\begin{equation}
\dot{x} = 
\begin{bmatrix}
0 & -0.363 & 0 \\
0.927 & -1 & 1 \\
0.063 & 0 & 0
\end{bmatrix} x +
\begin{bmatrix}
0.073 \\
0 \\
0
\end{bmatrix} v,
\quad
y = 
\begin{bmatrix}
1 & 0 & 0
\end{bmatrix} x
\label{eqn:linearized}
\end{equation}
where, the state vector is \( x = [p_{H_2},\ W_{H_2,\text{out}},\ q]^\top \), \( y = p_{H_2} \), \( v = P_{\text{in,mod}} \) is the applied power to the electrolyzer, which is the modified version of the input power \( r = P_{\text{in}} \). Here, \( x \), \( y \), \( v \), and \( r \) represent deviations from steady-state values in the linearized model, though this is not reflected in the notation for simplicity. Note that the resulting linear model has three states, whereas the nonlinear model has four. This is because changes in oxygen pressure do not affect the hydrogen pressure and are therefore unobservable from the perspective of the linearization.

In order to implement the PG for pressure management, the linear system model in \eqref{eqn:linearized} is first discretized with a sampling time of \( T_s = 0.1\,\text{s} \). Then, using \eqref{eqn:omega_e}, the set \( \Omega \subset \mathbb{R}^4 \) is characterized by 190 linear inequalities and determined offline. During real-time operation, at each sampling instant, the PG is applied using \eqref{eqn:v(t)}--\eqref{eqn:kappa} to enforce the hydrogen pressure constraint.

\begin{figure}
\begin{center}
\includegraphics[width=8.4cm]{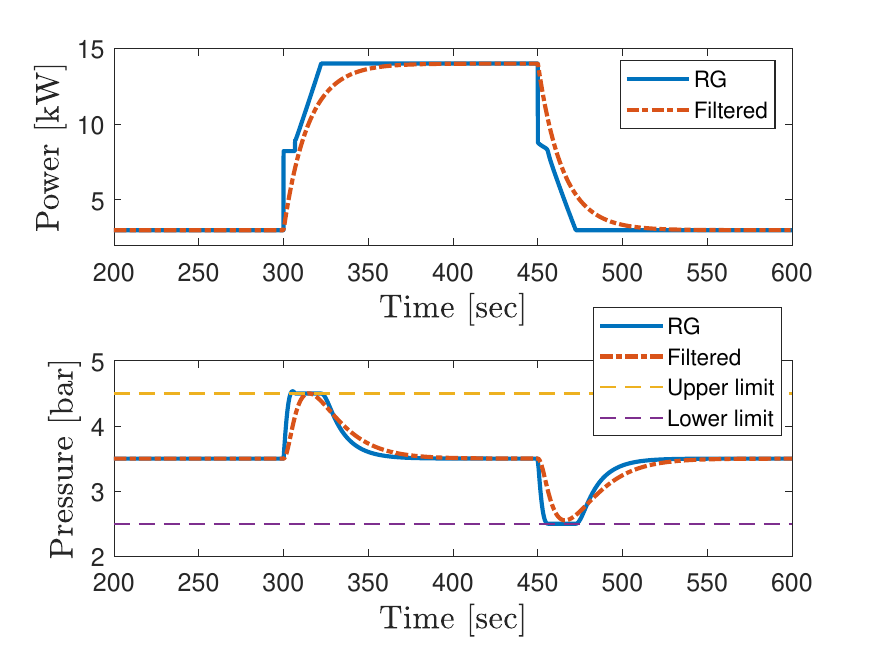}  
\caption{Comparison of power governor and conventional low-pass filter for pressure constraint management under
large step changes in input power.} 
\label{fig:pressure_L}
\end{center}
\end{figure}

To show the effectiveness of the proposed approach, electrolyzer with the PG is tested on a large step power, where the input power to the system steps from 3 to 14 kW. Fig.~\ref{fig:pressure_L} shows the responses of the hydrogen pressure and the applied input power to the system. In Fig.~\ref{fig:pressure_L}, the performance of the PG is also compared to the case where the input is low-pass filtered using a conventional first-order filter instead of an RG. For the design of such a filter, the time constant is tuned so that, in the worst-case scenario, that is, the largest step change in input power, the constraint is satisfied. The results in Fig.~\ref{fig:pressure_L} is when the time constant of the LPF is adjusted to  $\tau_F=14.5 s$. The simulation results confirm that the PG can maintain the  pressure within its constraint, with only a small violation occurring during the $11$ kW step-up in power input. This violation is due to the limited accuracy of predicting hydrogen pressure dynamics using a linear  model. However, the violation lasts no more than a few sampling periods and could potentially be eliminated using a disturbance bias approach, as presented in \cite{ayubirad2023simultaneous}, though this is not explored in this paper.

The simulation results indicate that, to avoid an upper-bound violation of hydrogen pressure during a step-up in input power, the PG applies less power to the electrolyzer than is available. 
From an energy management perspective, the excess power can be redirected to charge an auxiliary power buffer. Similarly, to avoid a lower-bound violation during a step-down in input power, more power is applied to the electrolyzer than is available. The power shortfall must thus be supplied by an auxiliary power source connected to the electrolyzer (e.g., the same power buffer). The results of the tracking performance, expressed in terms of mean squared error (MSE), along with hydrogen production during step-up and power saving during step-down for each method, are summarized in Table 1. As shown in Table 1, the PG outperforms the  LPF, improving power tracking error by 59.47\%, hydrogen production by 4.7\%, and auxiliary source energy consumption by 53.7\%.

\begin{figure} 
\begin{center}
\includegraphics[width=8.4cm]{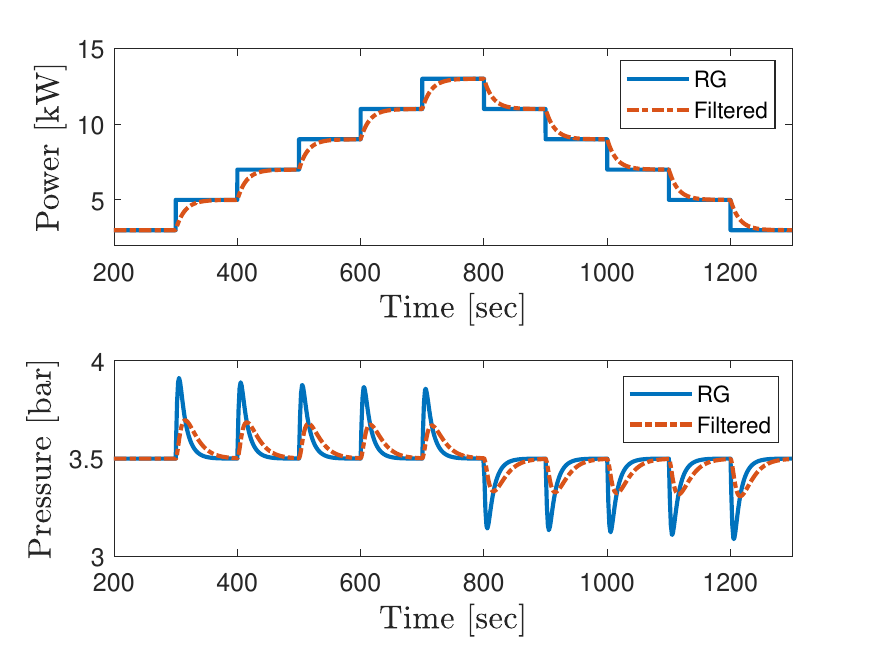} 
\vspace{-3mm}
\caption{Comparison of PG and conventional LPF for pressure constraint management for a sequence of small step changes in input power.} 
\label{fig:pressure_s}
\end{center}
\end{figure}

An attractive feature of the PG is its ability to slow down the system response only when active constraints are encountered. This is in contrast with the conventional LPF, which applies the same filtering indiscriminately to all input changes, even small ones that pose no risk of violating constraints. This behavior is illustrated in Fig.~\ref{fig:pressure_s},  where a sequence of small step changes in input power is introduced to the system. The results for tracking performance, hydrogen production during step-ups, and power saving during step-downs, as presented in Table~2, confirm the superior performance of the PG compared to the LPF when the introduced power changes pose no risk of constraint violations.
\begin{table}
\centering
{{Table 1.} {Performance Comparison of PG vs. LPF for large step changes in input power} \par}
\vspace{2mm}
\small
\begin{tabular}{|c|c|c|}
\hline
{Metric} & {PG} & {LPF}  \\ \hline
Power tracking MSE [kW] & 1.7947 & 4.4280  \\ \hline
H$_2$ production [Nm$^3$] & 0.0935 & 0.0893  \\ \hline
Auxiliary energy use [kWh] & 0.0205 & 0.0443 \\ \hline
\end{tabular}
\begin{tablenotes} 
\item \hspace{-1em} \textbf{Note}: Power tracking is evaluated over 200–600\,s. Hydrogen production is calculated during a step-up in input power (200–400\,s), and auxiliary energy use is measured during a step-down in input power (400–600\,s).
\end{tablenotes}
\label{tab:pg_lpf}
\end{table}
\begin{table}
\centering
{{Table 2.} {Performance Comparison of PG vs. LPF for a sequence of small step changes in input power} \par}
\vspace{2mm}
\small
\begin{tabular}{|c|c|c|}
\hline
{Metric} & {PG} & {LPF}  \\ \hline
Power tracking MSE [kW] & 0 & 0.2642  \\ \hline
H$_2$ production [Nm$^3$] & 0.2816 & 0.2736  \\ \hline
Auxiliary energy use [kWh] & 0 & 0.0403 \\ \hline
\end{tabular}
\begin{tablenotes}
\item \hspace{-1em} \textbf{Note}: Power tracking is evaluated over 200–1300\,s. Hydrogen production is calculated during the step-ups in input power (200–800\,s), and auxiliary energy use is measured during the step-downs in input power (800–1300\,s).
\end{tablenotes}
\label{tab:pg_lpf2}
\end{table}

\section{Conclusion}
\label{sec:conclusion}

This paper developed a constraint-aware power governor based on the Reference Governor (RG) framework to manage pressure dynamics in water electrolysis systems under varying power input. The objective was to enforce manufacturer-defined pressure limits while minimizing unnecessary modification of the electrolyzer input power. Unlike conventional low-pass filters, which apply the same filtering indiscriminately to all input changes, even small ones that pose no risk of violating constraints, the proposed RG-based approach modifies power as little as possible and only when active constraints are encountered. Simulation results showed improved pressure regulation, enhanced hydrogen production, and reduced auxiliary energy consumption. Future work will focus on real-time implementation and extending the strategy to incorporate hydrogen-to-oxygen diffusion constraint management, which is critical for maintaining high gas purity and safe operation of alkaline electrolyzers at low-load conditions.

\bibliography{ifacconf}             
                                                   






\appendix
\section{Parameters used in the model} 
\begin{table}[htbp]
\centering
{{Table A.1} \quad {Parameter list} \par}
\vspace{2mm}
\small
\begin{tabular}{ll}
\toprule
\text{Thermodynamic Constants} & \text{Value [Unit]} \\
\midrule
$R_{\mathrm{O_2}}$  & 259.8\;[J/(kg$\cdot$K)] \\
$R_{\mathrm{H_2}}$  & 4124.5\;[J/(kg$\cdot$K)] \\
$\gamma$            & 1.4\;[-] \\
$F$                 & 96485\;[C/mol] \\
\midrule
\text{Electrolyzer Parameters} & \\
\midrule
$V_{\mathrm{an}}$   & $2 \times 10^{-3}$\;[m$^3$] \\
$V_{\mathrm{ca}}$   & $1 \times 10^{-3}$\;[m$^3$] \\
$\tau_{\mathrm{ca}}$ & 1\;[s] \\
$C_D$               & 0.0124\;[-] \\
$A_T$               & $5 \times 10^{-5}$\;[m$^2$] \\
$n_{\mathrm{cell}}$ & 21\;[-] \\
$A_{\mathrm{cell}}$ & 0.25\;[m$^2$] \\
\midrule
\text{PI Gains}  \\
\midrule
$K_{p,ca}$ & $-0.93$\;[Nm$^3$/ (h$\cdot$bar)] \\
$K_i$ & $-0.06$\;[Nm$^3$/ (h$\cdot$bar$\cdot$s)] \\
\bottomrule
\end{tabular}
\label{tab:params}
\end{table}

\end{document}